\documentclass[conference]{IEEEtran}
\IEEEoverridecommandlockouts
\usepackage{cite}
\usepackage{amsmath,amssymb,amsfonts}
\usepackage{algorithmic}
\usepackage{graphicx}
\usepackage{textcomp}
\usepackage{xcolor}
\usepackage{multirow}

\def\BibTeX{{\rm B\kern-.05em{\sc i\kern-.025em b}\kern-.08em
    T\kern-.1667em\lower.7ex\hbox{E}\kern-.125emX}}
\begin{document}

\title{MRI Volume-Based Robust Brain Age Estimation Using Weight-Shared Spatial Attention in 3D CNNs}

\author{\IEEEauthorblockN{1\textsuperscript{st} Vamshi Krishna Kancharla}
\IEEEauthorblockA{\textit{Centre for Brain Research} \\
\textit{Indian Institute of Science}\\
Bangalore, India }
\and
\IEEEauthorblockN{2\textsuperscript{nd} Neelam Sinha}
\IEEEauthorblockA{\textit{Centre for Brain Research} \\
\textit{Indian Institute of Science}\\
Bangalore, India}
}

\maketitle

\begin{abstract}
Important applications of advancements in machine learning, are in the area of healthcare, more so for neurological disorder detection. A crucial step towards understanding the neurological status, is to estimate the brain age using structural MRI volumes, in order to measure its deviation from chronological age. Factors that contribute to brain age are best captured using a data-driven approach, such as deep learning. However, it places a huge demand on the availability of diverse datasets. In this work, we propose a robust brain age estimation paradigm that utilizes a 3D CNN model, by-passing the need for model-retraining across datasets. The proposed model consists of seven 3D CNN layers, with a shared spatial attention layer incorporated at each CNN layer followed by five dense layers. The novelty of the proposed method lies in the idea of spatial attention module, with shared weights across the CNN layers. This weight sharing ensures directed attention to specific brain regions, for localizing age-related features within the data, lending robustness. 

The proposed model, trained on ADNI dataset comprising 516 T1 weighted MRI volumes of healthy subjects, resulted in Mean Absolute Error (MAE) of 1.662 years, which is an improvement of 1.688 years over the state-of-the-art (SOTA) model, based on disjoint test samples from the same repository. To illustrate generalizability, the same pipeline was utilized on volumes from a publicly available source called OASIS3. From OASIS3, MRI volumes 890 healthy subjects were utilized resulting in MAE of 2.265 years. Due to diversity in acquisitions across multiple sites, races and genetic factors, traditional CNN models are not guaranteed to prioritize brain regions crucial for age estimation. In contrast, the proposed weight-shared spatial attention module, directs attention on specific regions, required for the estimation.
\end{abstract}


\section{Introduction}
Brain aging plays a pivotal role in cognitive decline, neurodegenerative diseases, and varying degrees of brain atrophy. Distinguishing between normal aging and pathological changes is an important challenge especially as similar brain regions are affected in both healthy aging and dementia. One helpful approach to identifying unusual changes in the brain is through brain-age assessment, which compares the condition of a person's brain to what would be expected at their age. With the world’s population growing and cases of neurodegenerative disorders becoming more common, it’s increasingly important to understand how aging impacts the brain and contributes to these conditions. Developing effective methods for identifying individuals at high risk, monitoring disease progression, and tailoring treatments accordingly is crucial.  By studying age-related changes in the brain, we can gain insights into how diseases develop and progress, with differences between brain age and chronological age indicating potential health risks at different stages of life. This highlights the importance of further research in brain age assessment.
In recent years there has been a significant increase in using Machine learning and Deep Learning methods across domains, that includes biomedical Imaging techniques. Deep Learning and computer vision has shown significant impact on biomedical problems that can solve complex problems, provided data. It  has been quite successful in various tasks involving brain age estimation, using MRI scans of the brain. These tasks come under a regression problem with supervised learning settings. Given a sample ``x'' from the data distribution we are trying to estimate ``y" by learning the function $f_w(x)$ with  learnable weights ($w$) over the training.

Brain age estimation models that use deep learning architectures can be categorized based on whether they use a slice-based or voxel-based(MRI volume) approach for modeling input data. Our current approach prioritizes the voxel-based method, which retains all information by considering the entire volume, unlike the slice-based approach.
In the slice-based approach, we use 2D image slices extracted from a 3D volume and train them using 2D CNN architectures, which can be either pre-existing models or newly proposed ones. Training 2D CNN models is relatively straightforward, but it results in the loss of important information and relationships between the slices. Conversely, in the voxel-based approach, we prioritize model size (parameters) over performance, expecting high effectiveness. 

In our paper, we introduce a robust weight shared spatial attention 3D CNN model with approximately 5.4 million parameters, achieving a low MAE compared to previous methods showed in Table 1, which represents the State-of-the-Art (SOTA). To the best of our knowledge, we are the \textit{first to employ a Weight Shared Spatial Attention model with 3D CNN}. We train our model on two distinct datasets - ADNI\cite{b15} and OASIS3\cite{b16} dataset. We evaluate the model's performance on an unseen dataset, demonstrating its consistent low MAE across these datasets, which shows the robustness of our model.

Attention mechanisms\cite{b8}, initially introduced in NLP also become popular in computer vision, drawing inspiration from how humans naturally identify important details in complex images. These mechanisms adjust feature weights dynamically mimicking how our eyes focus on relevant aspects of a scene. This capability is crucial for various visual tasks, from recognizing objects to understanding videos. In our proposed method, we delve into the significance of attention mechanisms and explore their applications in enhancing brain age estimation using Spatial Attention. 
\begin{figure*}[h]
\centering
\includegraphics[trim=0cm 4cm 0cm 4cm,clip,height=6cm,width=14cm]{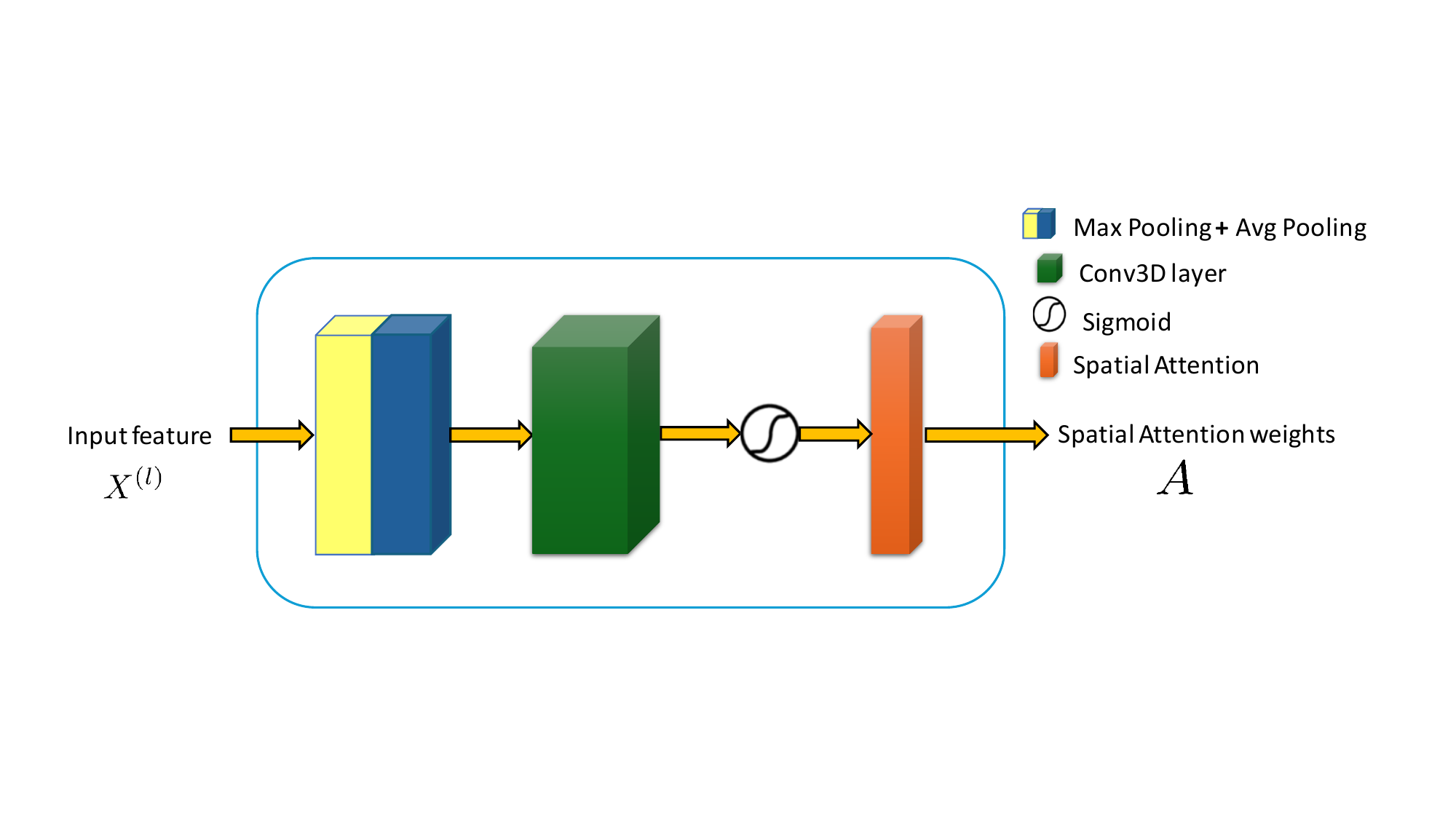}
\caption{Spatial Attention Module}
\label{fig:SP}
\end{figure*}
\section{RELATED WORK}
In the literature, several methods utilizing voxel-based analysis have been proposed.  Cole, James H., et al. \cite{b1} introduced a 3D CNN model that utilizes registered T1-weighted MRI image volumes segmented into white matter (WM) and gray matter (GM) for age prediction. They observed that GM volume outperformed WM, raw data, and their combination. The study also investigated the heritability of brain age estimation using GM, WM, and their combination, achieving a mean absolute error (MAE) of 4.16 years.  Bellantuono, Loredana, et al., \cite{b2} proposed complex networks employing a structural connectivity model. They segmented T1-weighted MRI scans into patches and measured the similarity between each pair of patches using Pearson correlation. These pairs of patches, along with their similarity scores, were then inputted into a deep learning model for regression, achieving an MAE of 2.19 years. Similarly,  Pardakhti, Nastaran,and Hedieh Sajedi., \cite{b3} proposed a 3D CNN model comprising 5 CNN layers followed by a fully connected layer. To enhance model accuracy, they combined 3D CNN features with two other models, namely support vector regression and Gaussian process regression, instead of using a fully connected layer, achieving an MAE of 5 years. Feng, Xinyang, et al., \cite{b4} introduced a 3D CNN model with a similar architecture used in Alzheimer’s disease classification. They employed a common CNN setup similar to the VGG classification model, involving stacking several blocks of convolutional layers and pooling layers, trained on a large and diverse dataset of structural brain MRIs, with an MAE of 4.06 years.

Kuo, Chen-Yuan, et al.,  \cite{b5} proposed an ensemble deep learning framework using 26 layers of the 3D ResNet. They demonstrated that combining various input feature sets and employing objective-specific functions through an ensemble deep learning framework improves the performance of brain age estimation methods, achieving an MAE of 3.33 years. Furthermore, He, Sheng, et al., \cite{b6} introduces a regression model based on deep relation learning, aiming to capture the relationships between pairs of input images. Four non-linear relations are considered simultaneously within a single neural network. This model performs two tasks: relation regression and feature extraction. EfficientNet is utilized for feature extraction, while a transformer is employed to model the relationships between image pairs. Evaluation is conducted on datasets containing subjects aged between 20 and 72 years, with MAE of 2.38 years, the lowest attained in the experiments.  

In our literature review, we concentrated specifically on voxel-based approaches for brain age estimation using deep learning techniques. While slice-based methods are another avenue of research in this field, our review did not delve deeply into them. Voxel-based methods utilize the entire brain volume, ensuring that no information is lost, unlike slice-based methods. This comprehensive representation enables voxel-based approaches to potentially yield more accurate and robust predictions of brain age. Therefore, our emphasis on voxel-based approaches in our review contributes to the current understanding and advancement of brain age estimation models.

\section{DATASET}

We have conducted our analyses on distinct datasets on Healthy subjects: ADNI\cite{b15} and OASIS3\cite{b16} comprising T1-weighted images. We performed preprocessing on the raw MRI scans before utilizing them for training our deep learning model. This preprocessing involved skull stripping and linear registration using FSL software. After preprocessing, our data shape consistently standardized to (91, 109, 91) across all datasets. From the ADNI\cite{b15} dataset, we selected 516 subjects, reserving 20\% for testing. Similarly, from OASIS3\cite{b16}, approximately 890 subjects were chosen with 20\% allocated for testing. 

We ensured an even distribution of subjects across age groups in each dataset to prevent bias in train and test. Specifically, the ADNI\cite{b15} dataset comprised subjects aged between 60 and 81 years, OASIS3\cite{b16} included subjects aged from 62 to 86 years.



\section{METHODOLOGY}
In this section, we'll discuss the proposed Weight Shared Spatial Attention in 3D CNN model, which utilizes spatial attention for brain age estimation.

\subsection{Spatial Attention Module}
Before delving into the proposed methodology, let's examine the Spatial Attention Module, as depicted in Figure~\ref{fig:SP}. Attention mechanism allocates more attention to important features in a given feature set. Woo, Sanghyun, et al. \cite{b7} proposed the Convolutional Block Attention Module, which integrates both channel attention and spatial attention, collectively termed as CBAM. The spatial attention mechanism operates by capturing the inter-spatial relationships of features. Unlike channel attention, which focuses on identifying important channels, spatial attention concentrates on determining the spatial location of informative parts within the input.

Inspired by this, for our brain age estimation task, we aim to leverage the spatial attention mechanism illustrated in Figure~\ref{fig:SP}. This mechanism dynamically allocates attention to important weights across different brain regions during model training. It involves three main operations: max pooling, average pooling, and convolution.  Initially, we conduct max pooling and average pooling along the channel axis in the feature maps to highlight significant regions. Next, we concatenate the results of these pooling operations to generate a comprehensive feature descriptor. On the concatenated feature descriptor, we apply a convolutional layer with sigmoid activation to produce a spatial attention map. This spatial attention map is element wise multiplied with input feature to get attention weighted feature.

\begin{figure*}[h]
\centering
\includegraphics[trim=0cm 1cm 0cm 0cm,clip,height=8cm,width=15cm]{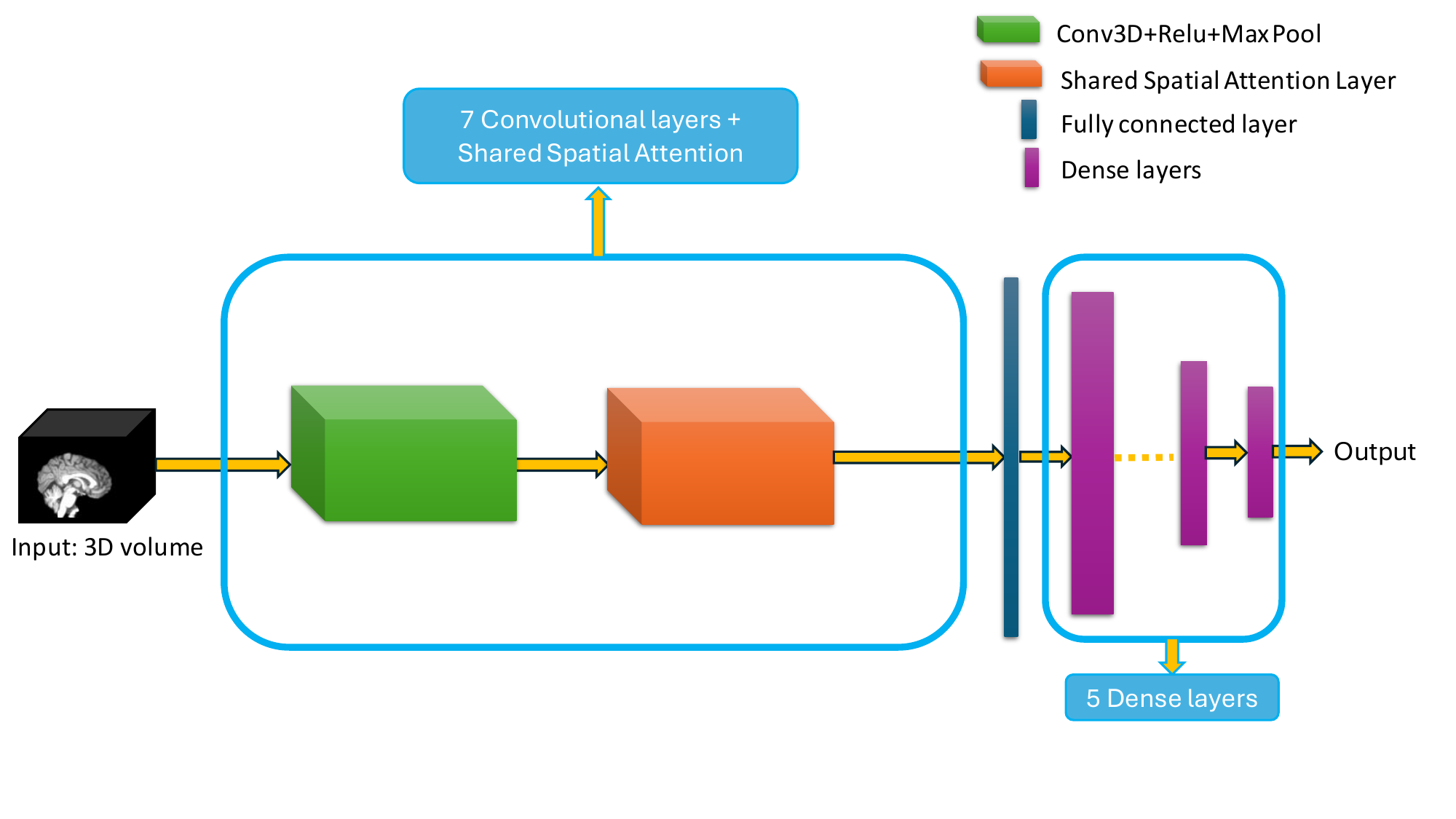}
\caption{Proposed framework for weight shared spatial attention in 3D CNNs}
\label{fig:BlockDiagram}

\end{figure*}

\subsection{Weight Shared Spatial Attention in 3D CNN model}

\begin{figure*}[h]
\centering
\includegraphics[trim=0cm 4cm 0cm 5cm,clip,height=4cm,width=10cm]{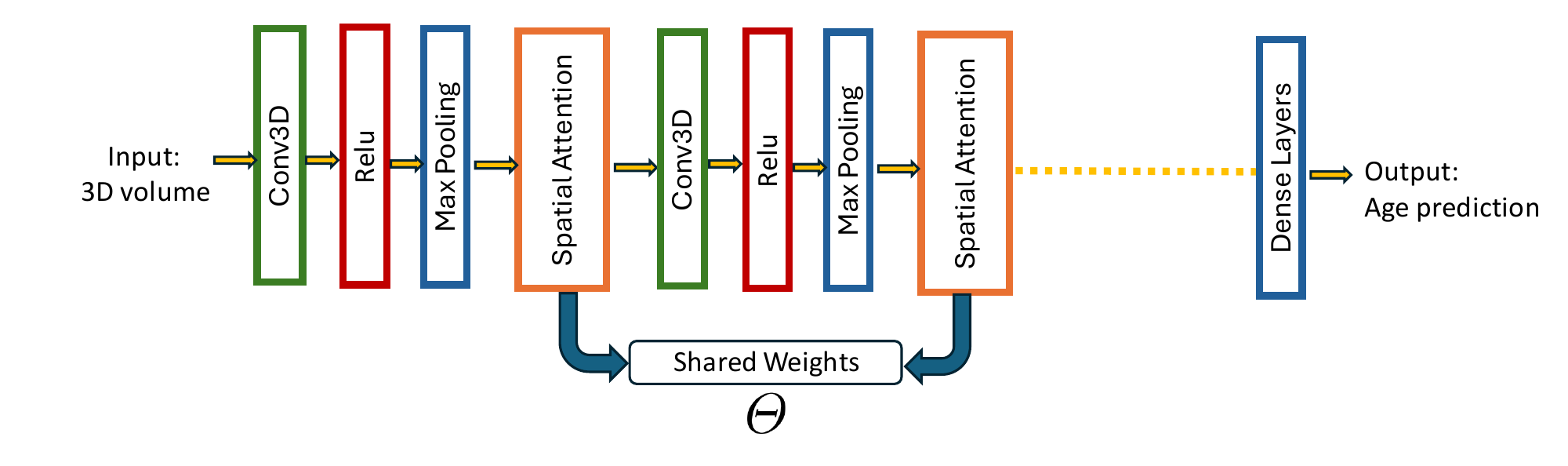}
\caption{Illustration of Weight Shared Spatial Attention across the convolution layers}
\label{fig:shared}
\end{figure*}

The proposed state-of-the-art Weight Shared Spatial Attention in 3D CNN model, as illustrated in Figure~\ref{fig:BlockDiagram}, comprises two main blocks. The first block consists of seven 3D convolutional layers, each of which is sequentially followed by weight shared spatial attention layers across all CNN layers. Following these layers is a flattening layer, which is subsequently connected to five dense layers. The architecture utilizes convolutional layers, each accompanied by Rectified Linear Unit (ReLU) activation functions, and incorporates max-pooling operations. After each max-pooling operation, a spatial attention layer with weights shared, as depicted in Figure~\ref{fig:shared}, is applied. These layers progressively extract hierarchical features from the input MRI volume, allowing the model to learn intricate spatial patterns while focusing on relevant brain regions and suppressing irrelevant features.
In the context of brain age estimation using brain MRI volume data, weight shared spatial attention mechanisms prove advantageous for several reasons. Firstly, they enable consistent focus on informative brain regions across multiple network layers, crucial for identifying relevant brain regions undergoing age-related changes. Secondly, by directing attention to specific brain regions during feature extraction, these mechanisms aid in localizing age-related features within the data, thereby enhancing prediction accuracy. Thirdly, the utilization of a single spatial attention mechanism across multiple layers reduces parameter redundancy, enhancing network efficiency. Finally, these mechanisms promote generalization and robustness by learning spatial patterns consistent across diverse populations and variations in brain anatomy and imaging protocols, facilitating superior performance across different datasets and individuals.

The spatial attention mechanism can be represented as:

\begin{equation}
A = \text{sigmoid}\left(\text{Conv}\left(\left[\text{MaxPooling}\left(X^{(l)}\right), \text{AvgPooling}\left(X^{(l)}\right)\right]\right)\right)
\end{equation}

\begin{equation}
W^{(l)} = A(X^{(l)}; \Theta)
\end{equation}
Equation [2] represents the computation of shared spatial attention weights $W^{(l)}$ using the spatial attention mechanism $A$, where $X^{(l)}$ is the input feature from the CNN layer at layer $l$ and $\Theta$ represents the weights shared of the spatial attention mechanism, shown in Figure~\ref{fig:shared}

The weight shared spatial attention-modulated features at layer \(l\) can be computed as:
\begin{equation}
Y^{(l)} = W^{(l)} \odot X^{(l)}
\end{equation}
where the element-wise multiplication ($\odot$) is applied between ($W^{(l)}$) and $X^{(l)}$.

During backpropagation, the gradients with respect to the weight shared \( \Theta \) of the spatial attention mechanism are computed as:
\begin{equation}
\frac{\partial L}{\partial \Theta} = \sum_{l} \frac{\partial L}{\partial Y^{(l)}} \cdot \frac{\partial Y^{(l)}}{\partial W^{(l)}} \cdot \frac{\partial W^{(l)}}{\partial \Theta}
\end{equation}
where $L$ is the loss function. The gradients from each layer contribute to updating the shared weight, enabling the spatial attention mechanism to adapt to information provided by multiple layers. This weight shared spatial attention mechanism enhances the robustness of the CNN model across datasets for brain age estimation. To validate this, the model initially trained on the OASIS3 dataset. Then, without any transfer learning or additional training on the ADNI dataset, the trained model was tested on these unseen datasets (ADNI Dataset), resulted in MAE values of 3.86 years, which were comparable to the OASIS3 MAE of 2.265 years. This indicates that the model's performance remained robust across different datasets without the need for specific adaptation or retraining on those datasets. We also conducted experiments with our 3DCNN model utilizing separate attention mechanisms at each convolutional layer, excluding the weight shared spatial attention module. This resulted in a 0.56 MAE higher difference across the datasets. The proposed model's architecture comprises seven convolutional layers, each with a kernel size of (2, 2, 2), padding, and stride set to (1, 1, 1), and increasing output channels (12, 16, 32, 64, 128, 512, 1024). These layers facilitate feature extraction at various abstraction levels. Dropout regularization is applied after the last convolutional layers and also after the first dense layer to mitigate overfitting. Following the convolutional layers, the architecture integrates five dense layers (512, 128, 64, 12, 1), progressively reducing the feature dimensionality to produce a single output value for age prediction. ReLU activation functions are applied across the layers. The model is trained using the Adam optimizer with a learning rate of 0.0001 for 250 epochs.

\begin{figure*}[h]
\centering
\includegraphics[trim=0cm 4cm 0cm 1cm,clip,height=4cm,width=10cm]{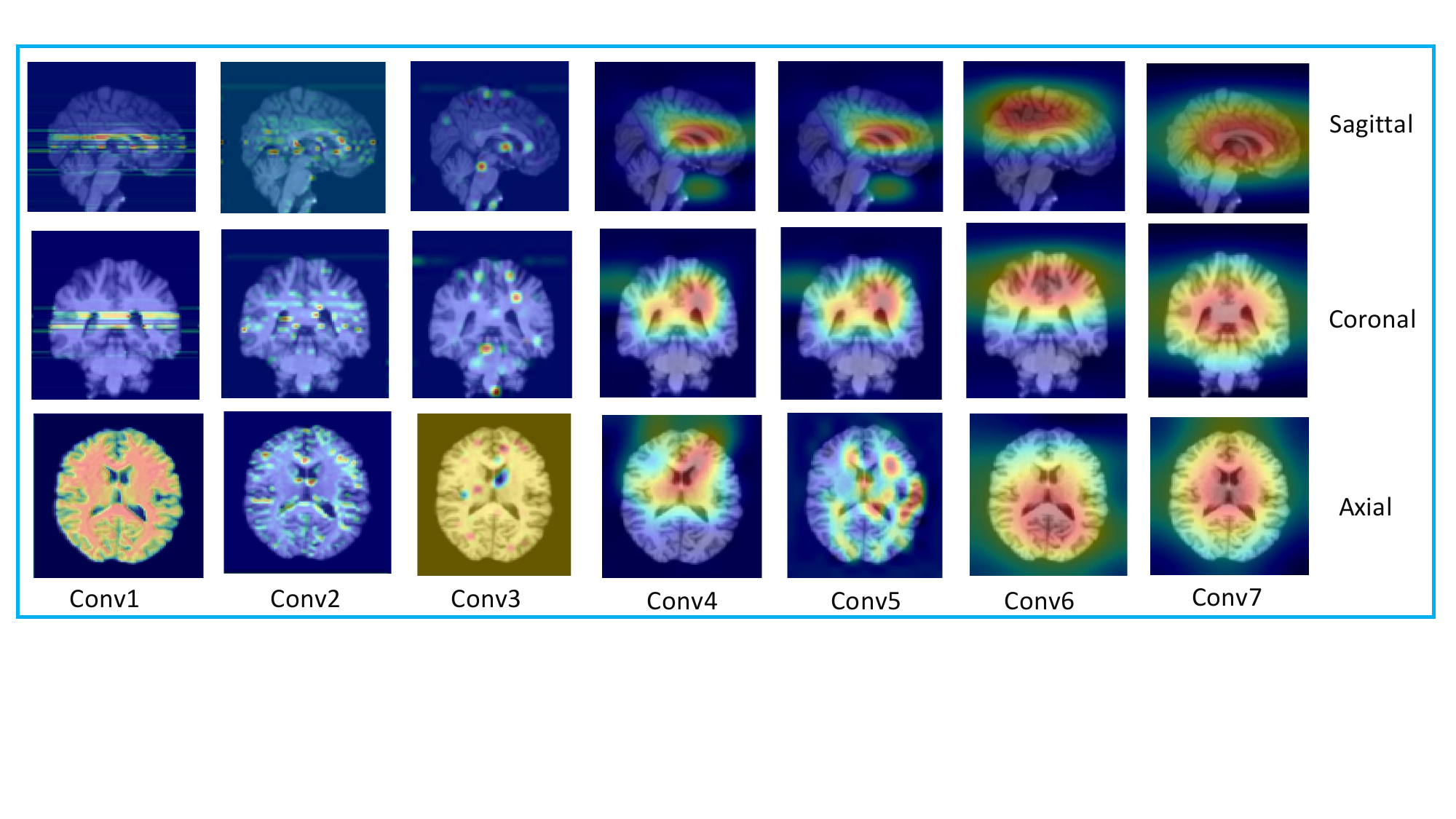}
\caption{Grad-CAM Interpretation across Convolution Layers in Sagittal, Coronal, and Axial Planes.}
\label{fig:Gradcam}
\end{figure*}


\begin{table*}[]
\centering
\begin{tabular}{|l|l|l|}
\hline
\textbf{Model}                  & \textbf{MAE}   & \textbf{Dataset}                                                                                       \\ \hline
SFCN\cite{b6}                    & 2.38           & \begin{tabular}[c]{@{}l@{}}TMGHBCH, NIH-PD, ABIDE-I, BGSP, \\ BeijingEN, IXI, DLBS, OASIS\end{tabular} \\ \hline
SFCN\cite{b11}                    & 3.85 ± 2.9     & T1-weighted MRI (CNNT1) and TOF MRA (CNNTOF)
                                                                   \\ \hline
3D CNN (ResNe)\cite{b12}         & 2.84 ± 0.5     & German National Cohort                                                                                      \\ \hline
Ensemble Deep Learning \cite{b13} & 3.33           & PAC2019                                                                                                \\ \hline
EfficientNet\cite{b14}            & 3.31           & NAC, IXI and ADNI                                                                                      \\ \hline
\multirow{2}{*}{\textbf{Ours}}    & \textbf{1.622} & \textbf{ADNI}                                                                                          \\ 
                                 & \textbf{2.265} & \textbf{OASIS3}                                                                                        \\ \hline
\end{tabular}
\caption{Comparison between our method's MAE and SOTA methods}
\end{table*}

\section{RESULTS AND DISCUSSION}
The obtained results show that the average MAE obtained across all 2 datasets is 
1.9435 years, which is comparable with reported SOTA techniques. However, when performance on each of the datasets is separately examined, it is seen that the performance on ADNI brain volumes is the best at MAE of 1.622 years, while the data from OASIS results in MAE of 2.265 years. 
A comparison of results obtained with those reported in literature is shown in Table 1. Besides, the deviation from ground truth is calculated using Root Mean Square Error (RMSE), obtaining values of 2.36, 3.03 across ADNI and OASIS3 datasets. Subsequently, the results are also interpreted with Grad cam\cite{b9} across convolution layers shown in Figure~\ref{fig:Gradcam}. Analysis of Grad-CAM reveals that the proposed  model predominantly focuses on the middle region of corpus callosum.

The proposed method has been applied on two datasets with varying resolutions. The brain volumes in OASIS dataset have poor spatial resolution, with grainy and blurred structures. This explains why they result in below-par performance, among the two datasets. On the other hand, images in ADNI dataset have higher spatial resolution with clear anatomical details. This leads to the best performance among the datasets considered. 
The proposed approach consistently outperforms many other approaches in literature. This is mainly attributed to the proposed spatial attention which is invariant across all the CNN layers. Popular works in literature have utilized layer-specific attention modules which lead to varied receptive fields focusing on different aspects of input. However, in our approach, we utilize single shared attention weights across all the layers. This leads to the focus on the same specific area across the brain volume. This mechanism ensures that the aspects of invariance in input brain volumes are considered in spite of diversity in the dataset.  The codes will soon be uploaded to the GitHub page.


\section{CONCLUSION}
In this paper we propose robust brain age estimation using a weight shared spatial attention module in 3D CNN. The input to this system are 3D structural MR Brain Volumes. The advantage of this approach is that the network is able to focus on those features that are invariant, inspite of the diversity in brain volumes among the different datasets. The performance of the proposed medthod as been illustrated on two different datasets which are publicly available ADNI and OASIS3 dataset. Peak performance, MAE of 1.622 years was obtained on 102 subjects of the ADNI dataset.

\end{document}